\documentclass[journal]{IEEEtran}
\usepackage{color}
\usepackage[dvipsnames]{xcolor}

\usepackage{cite}
\ifCLASSINFOpdf
  \usepackage[pdftex]{graphicx}
\else
  \usepackage[dvips]{graphicx}
\fi

\usepackage{multirow}
\ifCLASSOPTIONcompsoc
  \usepackage[caption=false,font=normalsize,labelfont=sf,textfont=sf]{subfig}
\else
  \usepackage[caption=false,font=footnotesize]{subfig}
\fi
\usepackage{fixltx2e}
\begin{document}
% \title{Bare Demo of IEEEtran.cls\\ for IEEE Journals}
\title{Optical Network Virtualisation using Multi-technology Monitoring and SDN-enabled Optical Transceiver}

% \author{Michael~Shell,~\IEEEmembership{Member,~IEEE,}
%         John~Doe,~\IEEEmembership{Fellow,~OSA,}
%         and~Jane~Doe,~\IEEEmembership{Life~Fellow,~IEEE}% <-this % stops a space
% \thanks{M. Shell was with the Department
% of Electrical and Computer Engineering, Georgia Institute of Technology, Atlanta,
% GA, 30332 USA e-mail: (see http://www.michaelshell.org/contact.html).}% <-this % stops a space
% \thanks{J. Doe and J. Doe are with Anonymous University.}% <-this % stops a space
% \thanks{Manuscript received April 19, 2005; revised August 26, 2015.}}
\author{Yanni~Ou, Matthew Davis, 
        Alejandro Aguado, Fanchao Meng, Reza Nejabati and Dimitra Simeonidou % <-this % stops a space
\thanks{Yanni Ou, Matthew Davis, Fanchao Meng, Reza Nejabati and Dimitra Simeonidou are with the High Performance Networks group in the Department of Electrical and Electronic Engineering, University of Bristol, BS8 1UB, UK, e-mail: yanni.ou@bristol.ac.uk.}% <-this % stops a space
\thanks{A. Aguado was with the High Performance Networks group in the Department of Electrical and Electronic Engineering, University of Bristol, BS8 1UB, UK, and now is with the Center for Computational Simulation, Universidad Politecnica de Madrid, 28660, Madrid, Spain.}% <-this % stops a space
% \thanks{Manuscript received xxxxx, 2017; revised xxxx, 2017.}}
}

\maketitle

\begin{abstract}

We introduce the real-time multi-technology transport layer monitoring to facilitate the coordinated virtualisation of optical and Ethernet networks supported by optical virtualise-able transceivers (V-BVT). A monitoring and network resource configuration scheme is proposed to include the hardware monitoring in both Ethernet and Optical layers. The scheme depicts the data and control interactions among multiple network layers under the software defined network (SDN) background, as well as the application that analyses the monitored data obtained from the database. We also present a re-configuration algorithm to adaptively modify the composition of virtual optical networks based on two criteria. The proposed monitoring scheme is experimentally demonstrated with OpenFlow (OF) extensions for a holistic (re-)configuration across both layers in Ethernet switches and V-BVTs.

\end{abstract}

% Note that keywords are not normally used for peer review papers.

% \fi
\IEEEpeerreviewmaketitle

%  --------------------------- introduction ----------------------------------------
\section{Introduction} \label{introduction}

% why optical virtualization
% why monitoring -> facilitate optical virtualizaiton
% why V-BVT -> hardware-level virtualization -> facilitate optical virtualizaion

\IEEEPARstart{F}{uture} Internet applications in the domains of science, business and domestic users \cite{Cisco2013CACiscoVisual,Vecchiola2009Highperformancecloud,Al-Fares2008scalablecommoditydata} are all observed to increasingly rely on a large number of powerful and often widely distributed hardware and software resources, as well as the network that interconnects them \cite{Develder2012Opticalnetworksgrid}. These resources have been growing exponentially (predicted by Moore's Law), and cloud services are currently the emerging trend to offer both distributed hardware and software delivering as a service on a global scale. The performance and availability of cloud services highly depend on the cloud physical infrastructure composed of data centre (DC) infrastructure, its inter- and intra-DC networking, as well as end connectivities to users. 
  
Optical networks that consist of novel technologies are considered the most promising network substrate under this condition. Optical network virtualisation is one of the key contributor \cite{Nejabati2011Optical,Duan2012surveyserviceoriented} to efficiently enable the combined management, control and optimisation of networking resources for Cloud service provisioning \cite{aguado2016dynamic}. Virtual optical networks (VON), composed of multiple virtual nodes interconnected by virtual links, are co-existing but isolated, sharing the same optical network substrate. Accordingly, the analogue constraints and impairments of the optical network substrate will have a great impact on VONs compositions and their performance.
 
Currently, physical layer impairment-aware models \cite{Cardillo2005Consideringtransmissionimpairments,Peng2013Impairmentawareoptical,Saradhi2009Physicallayerimpairment,carena2012modeling,poggiolini2012gnModel} have been studied under different network technologies, and some of them are introduced into optical virtualisation \cite{Peng2013Impairmentawareoptical,wang2017load,beyranvand2013quality,ou2015onlineoffline}. This method relies more on the pre-planned analytical estimation of impairments in the optical substrate (e.g., link nonlinearity calculation), while lacks the ability to adapt, e.g., it cannot compensate the undesirable and time-varying loss or excessive noise that causes a big degradation in the optical channel quality of transmission (QoT). Besides, due to diverse application types, traffic from these applications varies dynamically with time, which in turn greatly affects the allocation of virtual link and node resources to support the transmission of application traffic. 

Under this condition, it is important to introduce real-time monitoring across all the network layers as a key role in the virtualisation process, especially in tracking available resources in task scheduling. It should also include the status monitoring of already provisioned services and used physical resources \cite{Develder2012Opticalnetworksgrid,MonitorSDN2015,van2014opennetmon}.  
By obtaining and analysing monitored data, an up-to-date understanding of the network dynamic will be formed from different aspects of the network. This understanding will further affect the (re-)configuration of network resources in supporting both existing and new services to achieve IT/network elasticity, service-level agreement (SLA) requirements and QoT (e.g.,). Meanwhile, in combination with SDN , the controller can realise and optimise such (re-)configurations to different users in an efficient manner\cite{simeonidou2013software,van2014opennetmon}. This is due to the intrinsic characteristics of SDN, such as the separated data and control planes, a centralised management, global network view, and data plane open interfaces \cite{amaya2014software,kreutz2015software}. 

Therefore, we propose a multi-technology monitoring scheme enabled by SDN to obtain up-to-date characteristics of the optical transport layer, including optical link QoT and link spectrum utilisation, as well as the Ethernet transport layer, e.g., Ethernet traffic data rate, packet size and deep packet inspection (DPI). The previous study \cite{Multi-tehcMonitoring2016OFC} mainly introduces an algorithm that takes the monitored optical network status and the VON traffic as its inputs. However, it did not address the source and capturing ways of the monitored information, the communication between the monitored information and the algorithm, as well as how the information is related to the control plane. Here in this work we address these issues in the following four aspects.

First, a complete monitoring scheme is presented that covers different network layers, including the interaction among the monitoring and (re-)configuration in the data plane, the SDN controller in the control plane, and the management/decision making in the applications layer.
Meanwhile, in this scheme, virtualisation of optical transceivers (V-BVT) \cite{ou2016demonstration} is also employed in the data plane to introduce a device-level (Layer-1) virtualisation into the optical network. It represents the physical transceiver to the control plane as an abstracted software object, enabling the on-demand creation of virtual transceivers that generates/terminates one or multiple virtual links within the VONs. Therefore, it can offer a fine partition of the physical resources and guarantee complete isolation of applications \cite{vecchiola2009high}, and accordingly offer further enhanced flexibility and efficiency to support VONs. Based on these functionalities, the proposed multi-technology real-time monitoring scheme aims to further facilitate the coordinated virtualisation of the packet transport network and optical transport network, in order to achieve a holistic optimisation in the optical layer and the configuration in the Ethernet layer. 
Furthermore, we discuss the device types that can be used for Ethernet or optical monitoring and how can they be enabled by SDN using OF extensions. The way or protocol of storing and retrieving the monitored data are also discussed from the algorithm's perspective. The potential of achieving multi-level monitoring in a feasible manner using SDN can be achieved. Second, we present a DC network use case of the proposed monitoring scheme to show the potential performance improvement considering resources allocation. Third, We elaborate the algorithm details on how it will perform corresponding to the interactions among different blocks from the proposed monitoring scheme. Finally, we add the results from the network interfacing card (NIC) monitoring that supports the DPI, showing the capture of Ethernet traffic that can be carried by the VON requests. Under the condition that a large variation of these traffic exists, we then can use this captured information to decide how to make the aggregation decisions at the OF-enabled Ethernet switch to achieve the optimised resource allocation. 

The paper is constructed as follows. Section II proposes the principle of a multi-technology transport layer monitoring scheme for a general SDN-based network environment that interconnects remote DCs and users. The scheme also elaborates on the employment of a virtualisation strategy on top of the SDN-enabled control plane to facilitate the optimisation of coordinated virtualisation using the monitored data and the proposed V-BVT. In Section III, the principle and detailed logic of the virtualisation strategy under the monitoring scheme are described. Two network scenarios are described corresponding to the interactions among the blocks from the proposed scheme. Section IV experimentally demonstrates the proposed scheme for several specific monitoring use cases, showing the re-configuration of Ethernet layer resources and the optimisation of V-BVT resource allocation in the optical layer for QoT maintenance. Finally, Section V concludes the paper.

\section{Optimisation of Optical Virtualisation using Monitoring Scheme and V-BVT} \label{sec2:scheme}

\begin{figure}[t!]
\centering
% \vspace*{-25pt}
\includegraphics[width=0.49\textwidth]{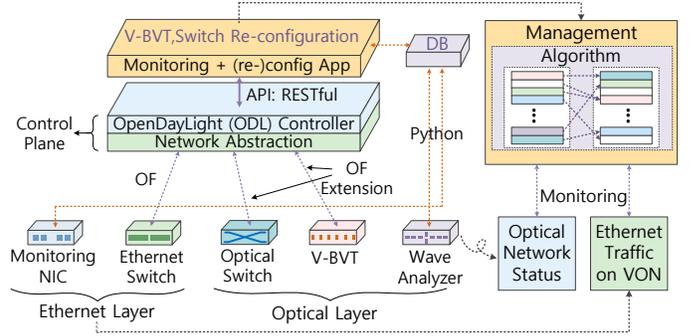}
% \vspace*{-40pt}
\caption{The application for V-BVT resource allocation and monitoring, DB: database, OF: OpenFlow.}
\label{fig:scheme}
\end{figure}

A high-level monitoring and network resource configuration scheme is proposed and described in Fig. \ref{fig:scheme}. At the bottom of the figure, Ethernet and optical layer hardware components are represented separately, which are all managed by an application based on top of a centralized controller. Some of these components are directly managed by the control plane that generally consists of a controller, i.e., the OpenDayLight Lithium (ODL) controller and the network abstraction layer. Components in different network layers communicate with the control plane using slightly different protocols. Ethernet switches in Ethernet layer are usually controlled using the standard OF protocol. In optical layer, optical transponders and optical switches require OF protocol extensions to be centrally managed. Following the virtualisation procedure, physical features of these hardware devices are firstly abstracted into the control plane, covering the range of bandwidth, port rate, power and other characteristics. The controller stores the information from these devices, and can also be queried via a representational state transfer (REST) application programming interface (API) for the usage of other applications.

The OF extension for our experiment is the extension of the V-BVT, which is equivalent to the extension of the WSS in our architecture. A dedicated OF agent is implemented for the WSS, containing a specific control protocol supported by the given optical devices. On top of it, a technology specific mapping function is implemented to translate the device data structure and protocol into the OF style, performing a set of actions such as wavelength, filtering and ports configuration. At the northbound side, each OF agent implements the extended OF protocol. For the SDN controller, the services abstraction layer was extended to record the support wavelength and supported spectrum range. To properly configure these devices, the forwarding rules manger was also extended to construct the required configuration information, e.g., central frequency, bandwidth, out-put port for the WSS together with match and label. More details of the agent, ODL and OF extensions have been described in the following a few public projects deliverables \cite{LightnessD4-4, LightnessD4-5,LightnessD5-2}. 

Other components, including the wave analyser in the optical layer and the network interfacing card (NIC) in the Ethernet layer that supports the DPI function, directly communicate with a database that sits in the application layer. The monitoring function is performed using wave analysers for the optical layer, while NICs and Ethernet switches are used from the Ethernet layer.

Wave analysers monitor a number of parameters and regard them as the optical network status, including the channel utilization on each path within the network and optical signal to noise ratio (OSNR) values for each channel and each of the already established lightpaths. NICs that support DPI will update the Ethernet layer monitoring parameters in the database which will be regarded as VON traffic. These parameters include Ethernet packet size, real-time data rate, packet MAC address and even deep-packet information (Layer 4 and upper). The Ethernet switch will report to the controller, providing real-time Ethernet traffic data rate on a per-port basis. This information is regarded as Ethernet traffic information carried on VONs in this scheme, and the polling frequency can be customised in the SDN controller. All the monitored information obtained from these devices is updated and stored to a database, and can be queried by applications.

For our experimental usage in Section IV, the proposed V-BVT is employed as an example of optical transponders, while optical switches used include fibre switches and wavelength selective switches (WSS). As well as the V-BVT being able to support virtualisation, its inclusion into the monitoring scheme enables efficient real-time response to variations in network. This is mainly due to the feature that the V-BVT is SDN-enabled, and its architecture allows the (re-)configuration of hardware resources in a flexible manner \cite{ou2016demonstration}.

On top of the controller, the monitoring and (re-)configuration application contains both monitoring and management functions. The monitoring function can periodically fetch both up-to-date Ethernet and optical monitoring information from the database and send it to the management block which contains a virtualisation strategy that executes the algorithm for hardware resource optimisation. For experimental demonstration, the algorithm is simplified so that the management block acts as a non-injective and non-surjective function, where multiple conditions may have the same action (and actions with non-active conditions). However, more realistically, the algorithm will cover more conditions when optimising the VON accommodation based on the given network and hardware resources. The analysis of the optimisation is beyond the scope of this paper and will be discussed in a separate one, where modulator utilization, modulator types and traffic conditions will be analysed. 

Any variations of information sent by the monitoring scheme will affect the V-BVT resource selection in creating virtual transceivers, e.g., modulation format and baud rate, the subcarriers central frequency and number, etc. Accordingly, when a variation is detected by the monitoring system, the management block will activate an action with a set of configurations, such as a change of optical channel selection and aggregation methods in the Ethernet layer. As explained in \cite{ou2016demonstration}, V-BVTs are placed at the edge of the optical network and each contains a local infrastructures pool, i.e., optical subcarriers pool and optical modulators pool. It can create multiple virtual transceivers based on the requirement of VON demands, the availability of its local infrastructure pool, and the optical network status.

\begin{figure}[t!]
\centering
% \vspace*{-10pt}
\includegraphics[width=0.43\textwidth]{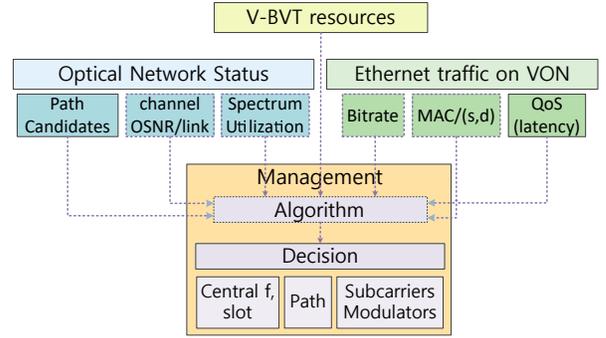}
% \vspace*{-10pt}
\caption{Inputs and outputs of management block in brief.}
\label{fig:Algoirthm_manage block}
\end{figure}
 
Fig. \ref{fig:Algoirthm_manage block} illustrates a generic logic of the virtualisation algorithm within the management block that adopts V-BVT and multi-technology monitoring. The detailed logic of the algorithm itself will be demonstrated in Section \ref{sec3:algorithm}. There are three inputs of the algorithm in the management block, including V-BVT resources, optical network status and the Ethernet traffic on VON. V-BVT resources contains the modulator and subcarriers resource pool, including the details of modulation types, modulator baud rates, the available number of each type, required OSNR for a given modulation type in a given baud rate, available subcarriers number, and central frequency of each available subcarrier \cite{ou2016demonstration}. The optical network status consists of pre-calculated path candidates between given source and destination pairs, channel OSNR per link based on a given transmitted data rate, latency and available spectrum slots of each link based on flex-grid from international telecommunication union (ITU). The last three parameters are updated by real-time optical monitoring techniques. Ethernet traffic on VONs consists of the real-time packet data rate, the packet MAC address and latency requirements. Path candidates and channel OSNR per link are updated by the Ethernet monitoring. The latency are assumed to be known by the algorithm but can further be gained by using deep-packet monitoring in the Ethernet layer using NICs that support DPI.

The management block outputs the algorithm decision with the objective of accommodating for the maximum number of incoming VON requests. The decision covers aspects from the three inputs. From the Ethernet perspective, the aggregation of services is decided, including how many services to aggregate into one and which service should be chosen for a specific aggregation. From the optical network perspective, the selection contains optical path selection, number and frequencies of spectrum slots for this path, as well as central frequency of this channel for this path. The selection in V-BVT resources consists of subcarrier central frequency, baud rate and modulation format of the modulator. The decision will be sent to the ODL controller through the RESTful API, and the controller will re-configure Ethernet Switches, WSSs and V-BVT according to the decision it received through OF and extended OF respectively. 

\begin{figure}[h!]
\centering
%\vspace*{-35pt}
\includegraphics[width=0.44\textwidth]{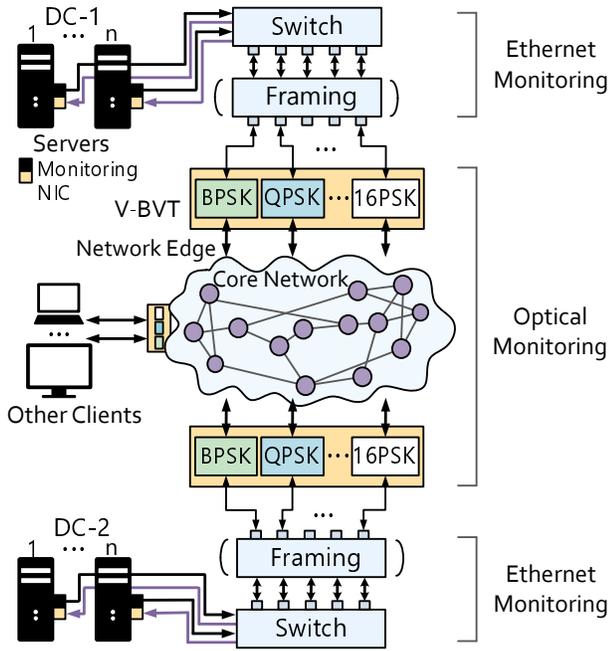}
%\vspace*{-45pt}
\caption{Applying Ethernet and Optical monitoring scheme in the scenario of core network and inter-DC connections.}
\label{fig:scheme in network}
\end{figure}

Fig. \ref{fig:scheme in network} shows an example of applying the proposed monitoring scheme in the optical core network and the inter-DC connections scenario. Traffic generated by servers in DC-1 is monitored by the NICs that support deep-packet inspection before entering in the aggregation switch, including MAC address and real-time data rate. The aggregation switch will route this traffic by reconfiguring the switch output ports. Ethernet traffic aggregation can be performed within the switch based on the monitored Ethernet information from the NICs. If traffic with the same destination MAC address enters from different ports, and if their real-time data rate is smaller than the maximum data rate of the output ports, they are considered for aggregation. Where port capacity allows, multiple input ports traffic will be aggregated onto a single output port.

In the framing block shown in Fig. \ref{fig:scheme in network}, Ethernet traffic from the aggregation switch is `re-framed' into either optical transport network (OTN) or customer-defined frames to suit the core network transmission. This can be performed by either an OTN switch or a customer-defined FPGA. After the framing, traffic frames with different data rate and quality of service (QoS) requirements will be modulated by the proposed V-BVT placed at the edge of the core network. It can accommodate different incoming traffic by selecting optical subcarrier frequency and modulation formats that have respective baud rates. Such decision is based on the syntactical analysis of V-BVT hardware resources (i.e., subcarriers pool and modulators pool), network status (i.e., link spectrum utilization, lightpath impairments and path candidates), together with required traffic QoS (e.g., latency), data rate and guarantee of QoT. Based on the decision, the V-BVT will modulate and transmit this traffic across the core network to its destination. When reaching the other edges of the network, traffic will be offloaded to customer sites, which can be another DC or other types of clients. 

% ------------------------------------ section 3 -------------------------------------
\section{Virtualisation Algorithm with Ethernet and Optical Monitoring Scheme} \label{sec3:algorithm}

\begin{figure}[h!]
\centering
% \vspace*{0pt}
\includegraphics[width=0.26\textwidth]{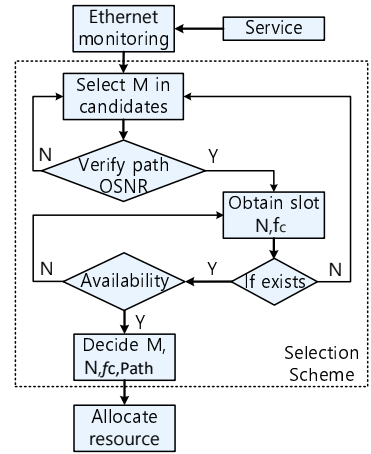}
% \vspace*{-10pt}
\caption{The application for the scenario of accommodating new services. $M$: modulation format types, $N$: number of contiguous spectrum slots, and $f_c$: central frequency}
\label{fig:Scenario_newservice}
\end{figure}

\begin{figure}[h!]
\centering
% \vspace*{-15pt}
\includegraphics[width=0.33\textwidth]{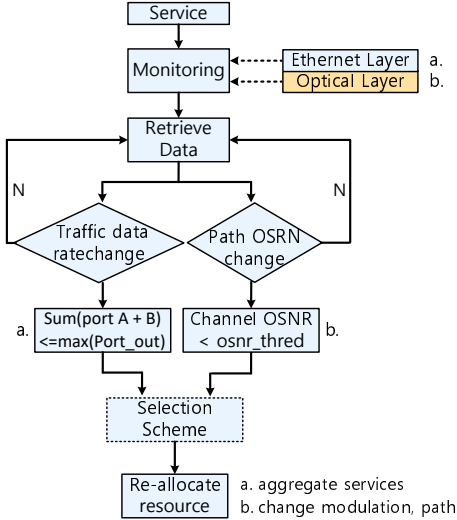}
% \vspace*{-10pt}
\caption{The application for the scenario of re-configuration by monitoring Ethernet and optical layers.}
\label{fig:Scenario_reconfig}
\end{figure}

Fig. \ref{fig:Scenario_newservice} and \ref{fig:Scenario_reconfig} show details of the management block used in the experimental demonstration. The first demonstration shown in Fig. \ref{fig:Scenario_newservice} accommodates a new incoming service based on its QoS requirement and a guaranteed QoT. When a new service is requested from the clients, e.g., VM transfer, through monitoring, the requirements of this request can be retrieved, including its data rate and source/destination pair (i.e., MAC address for Ethernet packet). This information (updated and stored in the database) will be retrieved by the algorithm selection scheme, in order to decide and verify if there are any Ethernet and optical resources available to accommodate the service. The audibility of these resources covers the areas of (i) Ethernet switching ports and the maximum ports capacity, (ii) the number of available contiguous spectrum slots $N$ to accommodate a given bandwidth, various modulator format types $M$ with different baud rates to meet the QoT requirements and decide the bandwidth, as well as the central frequency $f_c$ that can be used based on the ITU-T G.694.1 flex-grid standard for a given $N$ number of contiguous spectrum slots, (iii) network candidate paths and their spectrum utilisation between source and destination pair to meet the QoS requirements, etc.

The option will be selected if it can meet all the service requirements. Accordingly, configurations of physical devices, i.e., Ethernet switch and V-BVT, will be completed through the orders from the ODL enabled control plane, and a new service is then provisioned.

Fig. \ref{fig:Scenario_reconfig} indicates another scenario, where replanning of existing services will be triggered when two conditions happen: (a) a change in the Ethernet data rate of existing services, and (b) the change of channel OSNRs along the lightpath that is provided to existing services. 

In condition (a), the actual change of Ethernet traffic of one service is obtained by the Ethernet monitoring scheme using the NIC and Ethernet switch ports monitoring. The monitoring at these two devices covers two aspects of traffic. The SolarFlare NIC can monitor the traffic of different applications from different MAC addresses within the total amount of traffic, while the Ethernet switch can only monitor the total amount but not the higher layer packet information. When the real-time data rate of each service reduces, the summation data rate of $n$ services (which target the same MAC address, and $n$ represents A and B in the figure as an example) will be verified. If the summation is lower than the maximum Ethernet ports capacity, traffic from $n$ services is considered for aggregation and will be accommodated using one port instead of the original $n$ ports. V-BVT resources and network status will be re-verified as well, in order to guarantee the QoT and QoS of all the $n$ services. The details of the procedure are performed in the selection scheme that is introduced in Fig. \ref{fig:Scenario_newservice}. The procedure includes the re-verification/re-selection of subcarriers and modulation formats with respective baud rate in V-BVT resources, as well as candidate paths and spectrum slots in optical network. If both Ethernet and optical resources are available for the service aggregation, the aggregation option is selected. Optimised utilization of Ethernet switch resource and V-BVT hardware resources is achieved when accommodating these $n$ services. Re-configuration in both Ethernet and optical resources will be coordinated by the ODL controller. 

Similarly, for condition (b), we introduce optical layer monitoring to perform the re-creation of virtual transceivers from V-BVTs to support the same service. This condition will be triggered when the existing accommodation of a selected channel failed due to undesired optical network impairments. When the monitoring of an optical channel indicates an OSNR drop that will degrade the QoT of the existing service, another available channel that has enough spectrum slot will be decided together with a new modulation type that fits into this slot and QoT. 
In order to accommodate the service on the newly established optical channel, the central frequency of the spectrum will be configured within the V-BVT as well as the filtering width. Such establishment is also coordinated by the ODL controller using the extended OF protocol.

% ------------------------------------ section 4 -------------------------------------
\section{Experimental Demonstration of Monitoring and Optimisation Scheme} \label{sec4:experiment}

\begin{figure}[h!]
\centering
% \vspace*{-15pt}
\includegraphics[width=0.49\textwidth]{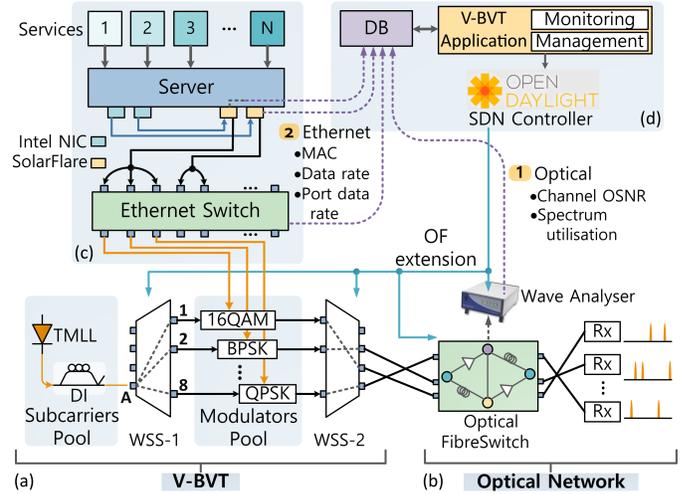}
% \vspace*{-10pt}
\caption{Experimental Platform: (a) V-BVT construction, (b) Arbitrary Network topology configuration using Polatis, (c) Ethernet layer configuration, (d) ODL enabled control plane with developed application.}
\label{fig:Monitoring_testbed}
\end{figure}

The experimental setup is represented in Fig. \ref{fig:Monitoring_testbed} for demonstrating the proposed monitoring scheme and the aforementioned scenarios. In corresponding with the architecture represented in Fig. \ref{fig:scheme}, the experimental realization of a V-BVT is displayed in inset (a) sitting in the lowest network layer; an optical fibre switch is displayed in inset (b) as part of the experimental optical network topology; Servers, NICs and Ethernet switch are also depicted inset (c), showing that the DC scenario and Ethernet layer sits upon the optical layer. Inset (d) shows the ODL controller enabled SDN control plane, in which a traffic engineering database is developed. On top of the control plane runs the V-BVT application that contains two monitoring blocks and the V-BVT virtualisation algorithm.

For inset (a), a similar setup for implementing V-BVTs is employed \cite{ou2016demonstration}. The subcarriers pool is settled using a tunable mode-lock laser (TMLL), and after applying the fibre delay interferometer (DI), around 25 optical subcarriers are selected in the spacing of 20 GHz and are sent to the input port A of the 4$\times$16 wavelength selective switch (WSS-1). The subcarriers pool includes a collection of modulators that can offer a range of modulation formats and baud rates, each of which are pre-connected to the 16 output ports of the WSS-1. The modulators consist of PM-16-QAM (10, 20, 28 GBd), BPSK (10, 40 GBd) and 10 GBd PM-QPSK. Different virtual transceivers are created after the selection of subcarriers and modulations. 

In inset (b), after sending the spectrum of the created virtual transceivers into the other 4$\times$16 WSS-2, the traffic that each virtual transceiver carrier can be directed onto same/different paths by selecting the output ports. The simplified optical network topology is settled using a 192$\times$192 optical fibre switch and coherent receiver is adopted to obtain the BER and constellations. The topology is composed of 4 nodes as a mesh network in the similar style of the one in \cite{ou2016demonstration} but using different fibre length. The links length used in this paper are three 50km, one 130km, and one 100km. Therefore, the shortest path between the same source and destination pair is 50km, and the longest is 150km.

Inset(c) demonstrates an example of Ethernet layer traffic (re-)directing and switch re-configuration based on real-time Ethernet traffic monitoring. An SDN enabled Pica 8 P-3922 Ethernet switch is employed. It can offer 10 GbE/40 GbE transmission with low latency, and support the running of Open-vSwitch (OVS) 2.0 to enable the OF interface. SolarFlare NICs (SFN5522) are also employed in this setup to represent the NICs that have DPI functionality for monitoring purpose. Each of the NICs contains two SFP+ interfaces, offering 10 GbE transmission and receiving.

The two and three input ports of the Ethernet switch come from VMs running in different servers to emulate the traffic from different applications. To monitor the variations of the traffic that emulate the time varying of different types of services or applications, the traffic is then generated slightly differently. This is mainly due to a limited number of VMs running in our lab servers. Therefore, we programmed in Python to generate PCAP (packet capture) file in one of our servers with different MAC addresses. We then sent the generated packages out over the NIC to emulate different application types running on that server. The traffic is generated as TCP/IP traffic. We let each of the traffic that represents one application type generates randomly but within the maximum capacity that a NIC interface rate should provide (10Gb/s SFP+). The outputs of the switch are pre-connected to the inputs of the modulator pool of V-BVT through an FPGA to emulate the ``framing" functions described in Fig. \ref{fig:scheme in network}. By selecting the output ports of the Ethernet switch, the indirect selection of modulation formats in the V-BVT is achieved for accommodating the input traffic from different VMs. However, in our experiment, the FPGA does not support the framing functionality, therefore, we used a compromising solution instead: the output of the Ethernet switch is sent to an FPGA, and the FPGA generates the corresponding 10 Gb/s or 40 Gb/s PRBS to feed into the optical modulators. Such configuration will not affect the network scenarios we tried to demonstrate. The output of the Ethernet switch will be accommodated by a selection of given different modulation formats, based on the output data rate, the required OSNR associated with the modulation levels, the available contiguous spectrum grids on the links, the central wavelengths, as well as the monitored real-time link OSNR condition. The decision making in this scenario is not seriously affected by the framing of the FPGA that can provide, but more about the data rate that the FPGA send to the modulators. Experiments with the FPGA that supports framing functionality are still worthy for future study.

The monitoring of the optical layer displayed in Fig. \ref{fig:Monitoring_testbed}inset (a) in the network is performed by applying the optical wave analyser (WA) that can offer 150 MHz resolution to obtain the OSNR values of each operational channel and the channel utilization on the link. The monitored information is updated into the traffic engineering database. 

The Ethernet monitoring is performed at the Ethernet switch input and output ports in inset (c) and the monitored real-time Ethernet traffic data rate variations are retrieved by the traffic monitor block in the application. The monitoring information from the database will be sent to the virtualisation algorithm to facilitate the decision. 

% ------------------------------------ section 5 -------------------------------------
\section{Results and Discussion} \label{sec5:Results}

\begin{figure}[t!]
\centering
\vspace*{0pt}
\includegraphics[width=0.43\textwidth]{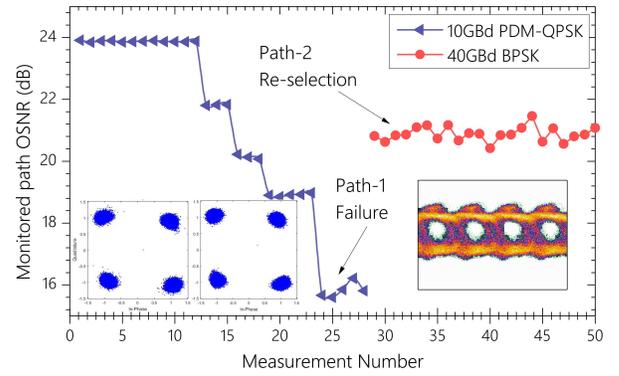}
% \vspace*{-10pt}
\caption{Real-time monitoring information: monitored optical link OSNR dropping with modulation format re-selection.}
\label{fig:link_OSNR}
\end{figure}

The results obtained from this experiment are from several aspects. Fig. \ref{fig:link_OSNR} shows the OSNR monitoring from the WA. We introduce external EDFA noise into the fibre to emulate the fibre link degradation. The original channel OSNR is of a high quality, around 24 dB. When the channel quality decreases gradually below the OSNR threshold of 15 dB, the failure alarm is triggered in the management , and the reconfiguration of modulation is activated as described in condition (b) of Fig. \ref{fig:Scenario_reconfig}. The algorithm considers the availability of the current link contiguous spectrum, central frequency $f_c$, required OSNR for transmitting this data rate using different modulation levels. It can provide a newly decided modulation, and a new channel or path can be selected from the V-BVT subcarriers pool. In this figure that represents one run of the algorithm, the newly provided path-2 has a shorter total distance and the monitored OSNR of the new channel is 20 dB. The modulation format 40GBd BPSK is decided to transmit the original data on the new path. The monitored OSNR before and after reconfiguration are shown, alongside a constellation and eye diagram representing the modulation formats before and after reconfiguration respectively.

\begin{figure}[h!]
	\centering
	% \vspace*{-15pt}
	\includegraphics[width=0.43\textwidth]{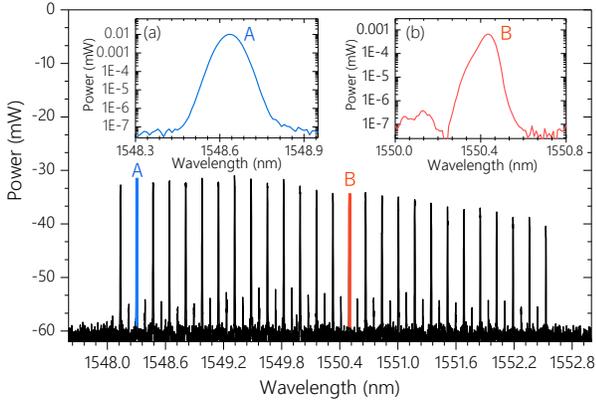}
	% \vspace*{-10pt}
	\caption{Real-time monitoring information: the original selected and re-configured subcarriers from subcarriers pool, (a) the original spectrum modulated by QPSK, and (b) the re-configured spectrum modulated by BPSK.}
	\label{fig:comb+spectrum}
\end{figure}

Fig. \ref{fig:comb+spectrum} represents the selection of different subcarriers from the subcarrier pool based on the monitoring of OSNR to reselect the path. For the first service, 40 Gb/s data traffic transmission using 10 GBd PM-QPSK is accommodated on one of the subcarriers within the pool at the wavelength of 1548.74 nm. This is depicted in the coloured tone A, and the spectrum of the created virtual transceiver is shown in Fig. \ref{fig:comb+spectrum} inset (a). 

% 

% \begin{figure}[h!]
%    \centering
%    \subfloat[Subcarriers pool]{
%    \includegraphics[width=0.31\textwidth]{comb_2_paint} \label{fig:comb_2}}
%
%   \subfloat[Spectrum on subcarrier A]{
%   \includegraphics[width=0.22\textwidth]{QPSK_spectra_paint} \label{fig:QPSK_spectra}}
%   %\quad
%   %\hfill
%   \subfloat[Spectrum on subcarrier B]{
%   \includegraphics[width=0.22\textwidth]{BPSK_spectra_paint} \label{fig:BPSK_spectra}}
%
%   \caption{Real-time monitoring information: (a) the original selected and re-configured subcarriers from subcarriers pool, (b) the original spectrum modulated by QPSK, and (c) the re-configured spectrum modulated by BPSK}
%   \label{fig:comb+spectrum}
% \end{figure}

Since the first service failed due to the OSNR below a threshold, the replacement service decided by the manager is automatically accommodated to another subcarrier on channel 1550.50 nm. The new modulation format based on this new channel condition are also decided and selected as 40 GBd BPSK for accommodating the same 40 Gb/s traffic transmission. The newly selected subcarrier is depicted in the coloured tone B, and the spectrum of the newly created virtual transceiver is shown in Fig. \ref{fig:comb+spectrum} inset (b).

\begin{figure}[t!]
\centering
\vspace*{-10pt}
\includegraphics[width=0.46\textwidth]{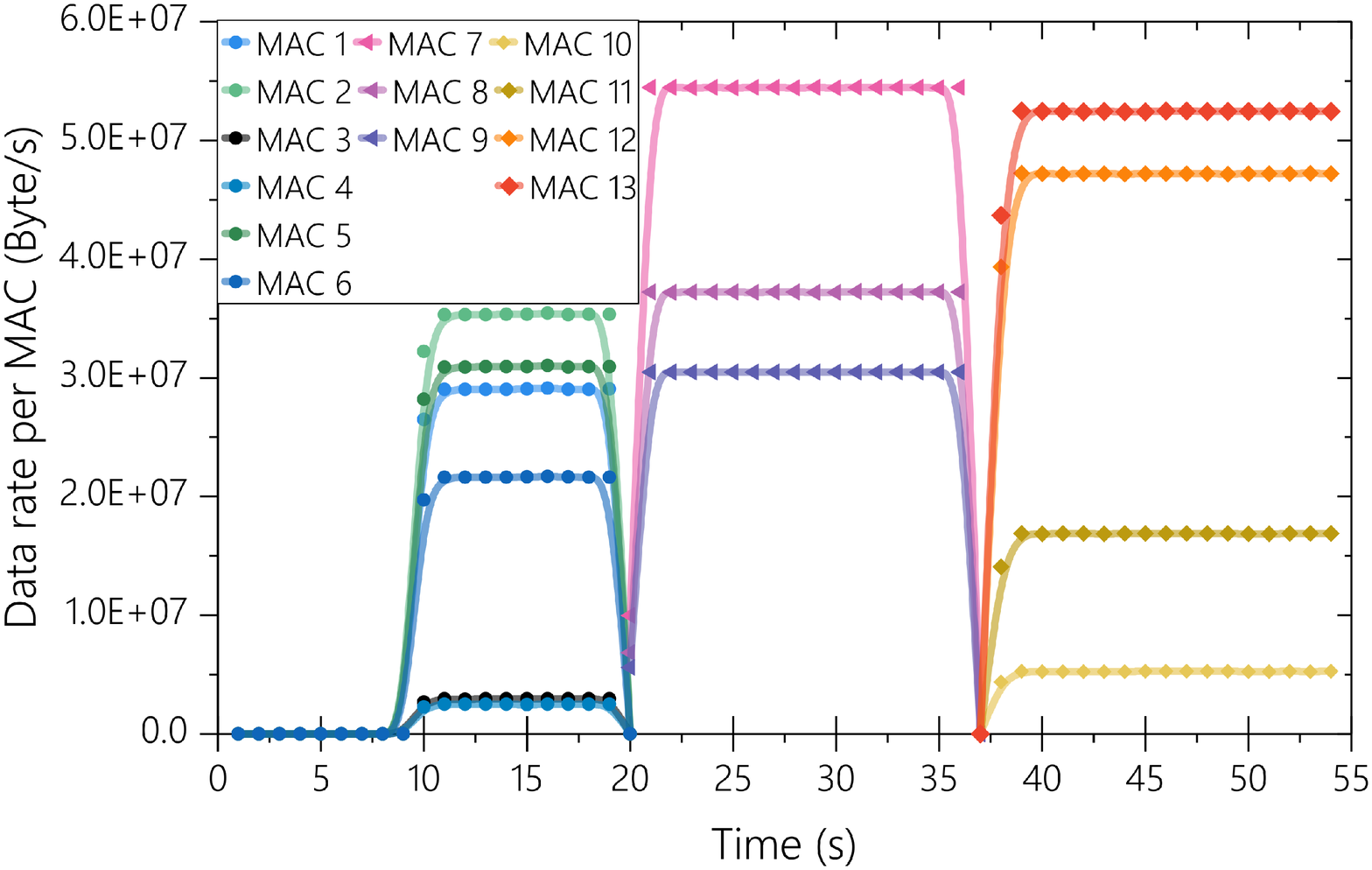}
\vspace*{-10pt}
\caption{Real-time monitoring: monitored Ethernet traffic data rate and MAC address variations from one port of one server along with time.}
\label{fig:NIC_monitoring}
\end{figure}   
Fig. \ref{fig:NIC_monitoring} reports the monitored dynamic from running services through one port of a SolarFlare NIC in the server, including the generation of different MAC addresses, and variations of Ethernet traffic data rate for each of the addresses along time. These MAC addresses and the traffic data rate variations are generated from the program PCAP file. The 13 MAC addresses aim to show the emulation of the varied customer access of different types of application types especially in a DC network scenario, including the frequent usage variations of each application type, its application duration and data rate. The variation of data rate comes from our program to generate random rate but within the maximum capacity that a NIC interface rate should provide.

Data is retrieved every second from the database, though the monitoring is in fact running at a finer time granularity. Service starts to run from the 8th second, randomly targeting six different MAC addresses (1-6), each of which has a respective data rate. Data rate summation of these six addresses is equal to 10 Gb/s, which is the maximum transmission rate of the SFP+ on the NIC. Services last for 10 seconds until the next three new services take their places, and last for the following 17 seconds. The newly targeted MAC addresses are captured by SolarFlare (7-9) as well as the newly updated data rate of each address. These services end by the 37th second, and another set of services start directly after and last after 55th second. Again, newly created MAC addresses (10-13) and their corresponding data rate are captured and updated on the plot.

%\begin{center}\footnotesize
\begin{table}[t!]
\caption{Monitored Ethernet MAC Addresses vs. Time}{}
\label{table: MAC table}
\centering
\begin{tabular}{ |c|c|l| }
  \hline
  \multicolumn{3}{ |c| }{Monitored MAC addresses} \\
  \hline
  Durations (s) & NO. & MAC address \\
  \hline
  \multirow{6}{*}{1-20} 
  & 1 & 59:53:83:2A:6:4C \\
  & 2 & 89:B4:C3:2:FE:8E  \\ 
  & 3 & 8A:26:8:A2:74:21 \\
  & 4 & A0:61:C:CA:EF:E1 \\
  & 5 & AB:25:1F:1D:AA:9B \\
  & 6 & B0:36:1E:2F:14:5A \\ %\cline{2-2}
  \hline \hline
  \multirow{3}{*}{21-37} 
  & 7 & A0:24:81:75:E5:DA \\
  & 8 & F2:74:62:F0:96:B9 \\
  & 9 & 25:40:F1:F0:E4:B8 \\ %\cline{2-2}
  \hline \hline
  \multirow{4}{*}{38-54} 
  & 10 & 36:B0:CD:68:3A:92 \\
  & 11 & 8D:72:35:B0:36:8F \\
  & 12 & B1:62:B8:C3:72:E1 \\
  & 13 & 11:6:89:CC:E:2A \\
  \hline
\end{tabular}
\end{table}
%\end{center}

In addition to Fig. \ref{fig:NIC_monitoring}, details of the obtained MAC addresses are listed in Table \ref{table: MAC table}. From 1 to 20 seconds, 6 MACs for the first group of services are captured, and likewise for 21 to 17 and 38 to 54 seconds.

\begin{figure}[h!]
\centering
\vspace*{-15pt}
\hspace*{-10pt}
\includegraphics[width=0.52\textwidth]{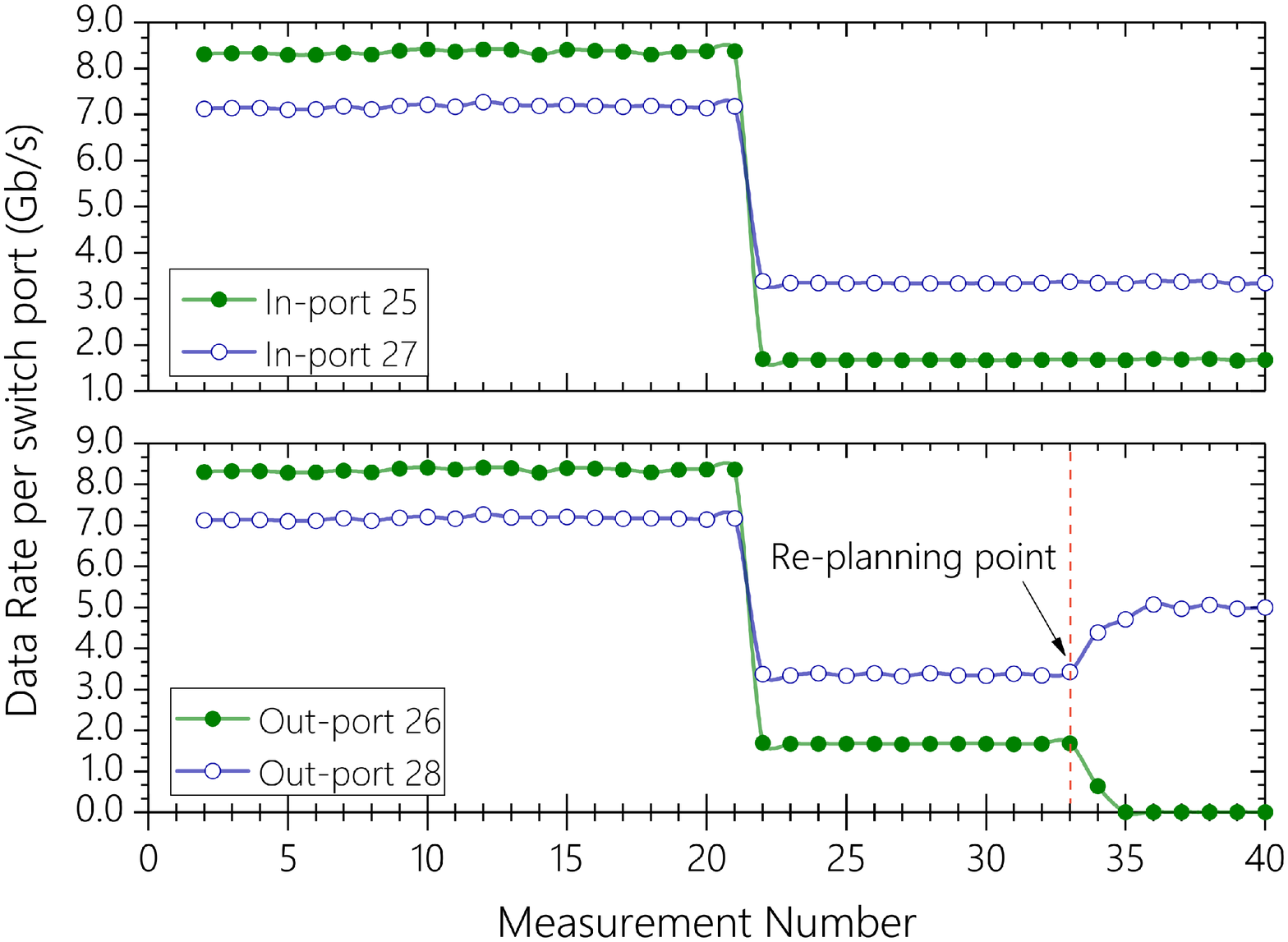}
\vspace*{-10pt}
\caption{Real-time monitoring: monitored Ethernet traffic at input and output ports of the switch, with traffic aggregation.}
\label{fig:Ethernet_Reconfig}
\end{figure}

Fig. \ref{fig:Ethernet_Reconfig} shows the monitoring of Ethernet traffic at the Ethernet switch. For the Pica 8 P-3922 Ethernet switch we used in this experiment, ports number 1-48 are for SFP+ 10GbE transmission and ports number 49-52 are for QSFP+ 40GbE transmission. The upper-plot illustrates the monitored traffic data rate variations from input ports 25 and 27 of the switch, while the lower-plot indicates the corresponding traffic of these two inputs from output ports 26 and 28 in the same switch. 

For the initial condition, which is the duration from 2 to 21 measurements, two new services are supported by two independent pairs of input and output ports of the switch. There are 8.6 Gb/s data rate from input port 25 and the same 8.6 Gb/s traffic from output port 26, as well as 7.2 Gb/s data rate from input port 27 and 7.2 Gb/s traffic from output port 28. The modulation format type to be used is decided by the algorithm described in the Fig. \ref{fig:Scenario_newservice} to accommodate these two new services. This decision is the equivalent of deciding which corresponding output port of the WSS should be enabled.

The output ports are indirectly decided by the algorithm from the Fig. \ref{fig:Scenario_newservice}, each of the services is to be accommodated using individual modulation format, baud rate and subcarrier from V-BVT resources, and will occupy individual bandwidth, path and channel (central frequency and number of grids) on the path in network. 

% do I need to add experiment on how to capture QoS in this case? otherwise selection scheme doesn't really work?

When condition changes, which starts from measurement 22, the monitoring scheme captures the traffic drop at input ports 25 and 27, from the original data rates becoming less than 2 Gb/s and 3.4 Gb/s respectively. This capture of traffic reduction triggers the condition (a) of the algorithm descried in the Fig. \ref{fig:Scenario_reconfig} to re-accommodate the two services, where the action of traffic aggregation is to be performed. Accordingly, both V-BVT resources (modulation format, baud rate and subcarrier) and network resources (path, channel central frequency and number of grids on the path) are to be re-verified. 

The new output port is selected as port 28, in this case based on the decided V-BVT and network resources, in order to accommodate the aggregated traffic. This can be seen from duration 33-40 measurements, where 33 is the re-planning point so the traffic in port 26 starts to transfer into port 28. Afterwards, port 26 is carrying no traffic, while monitoring on port 28 shows a rate of 5.4 Gb/s, accommodating the summation of the two dropped traffic from input ports 25 and 27. This allows the two services to be accommodated by only one switch output port, and V-BVT only needs to create one virtual transceiver to support the two services using a single modulator, a single subcarrier, a single path and the corresponding number of grids on that path. 

\begin{figure}[h!]
\centering
% \vspace*{-10pt}
\hspace*{-15pt}
\includegraphics[width=0.52\textwidth]{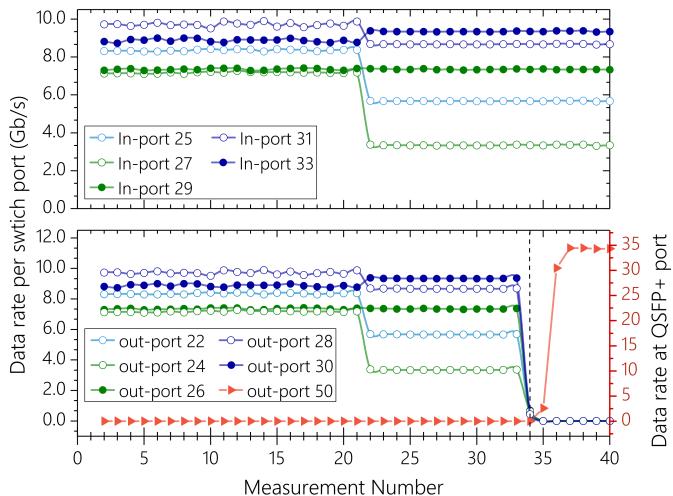}
\vspace*{-10pt}
\caption{Real-time monitoring: monitored Ethernet traffic at 5 input and output ports of the switch, with all 5 ports traffic aggregation at one QSFP+ output port.}
\label{fig:Ethernet_Reconfig_40G}
\end{figure}   

As well as monitoring the traffic at two pairs of input and output ports, in Fig. \ref{fig:Ethernet_Reconfig_40G}, the real-time traffic monitoring at five pairs of input and output ports is performed. Meanwhile, the newly aggregated traffic is in the level of 40 Gbps at the output port of the switch, transmitting the traffic through a QSFP+ instead of the SFP+ used in the Fig. \ref{fig:Ethernet_Reconfig}. From 0 to 20 measurements, the upper and lower sub-plots of Fig. \ref{fig:Ethernet_Reconfig_40G} both indicate the original traffic obtained from the 5 input SFP+ ports 25, 27, 29, 31 and 33, and 5 output SFP+ ports 22, 24, 26, 28 and 30. From the measurement of 21, the data rate of 5 input ports start to vary. Apart from the traffic in port 30 that slightly increases from 8.2 Gb/s to around 8.4 Gb/s, the data rate in the remaining ports all drop in various degrees. Again, the condition (a) of the algorithm described in Fig. \ref{fig:Scenario_reconfig} is triggered. With the sum of traffic throughput equalling around 35 Gb/s, it triggered the utilization of a QSFP+ at the switch output port 50 in Fig. \ref{fig:Ethernet_Reconfig_40G} to accommodate the transmission of the data that is currently transmitting at the 5 output ports 22, 24, 26, 28 and 30. In this case, the utilization of port sources at the switch is greatly reduced as the five services are accommodated only by one output port, which further reduces the V-BVT hardware and network resource occupation while serving the same data rate.

\begin{figure}[t!]
\centering
\vspace*{-13pt}
\includegraphics[width=0.42\textwidth]{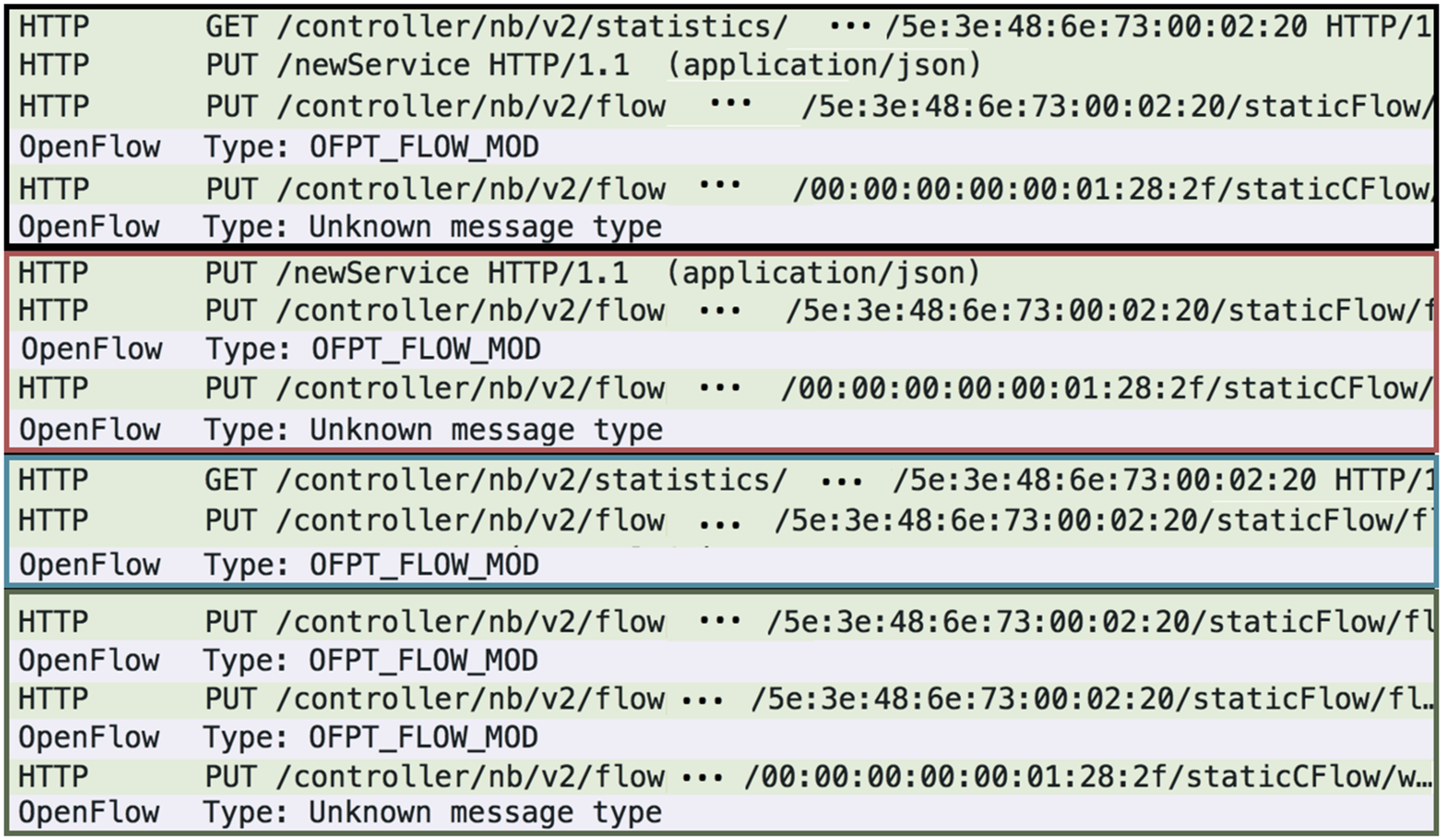}
\vspace*{-10pt}
\caption{Real-time monitoring information: flow messages for the configuration of the Ethernet switch.}
\label{fig:OF_Ethernet}
\end{figure}

\begin{figure}[t!]
\centering
\vspace*{0pt}
\includegraphics[width=0.45\textwidth]{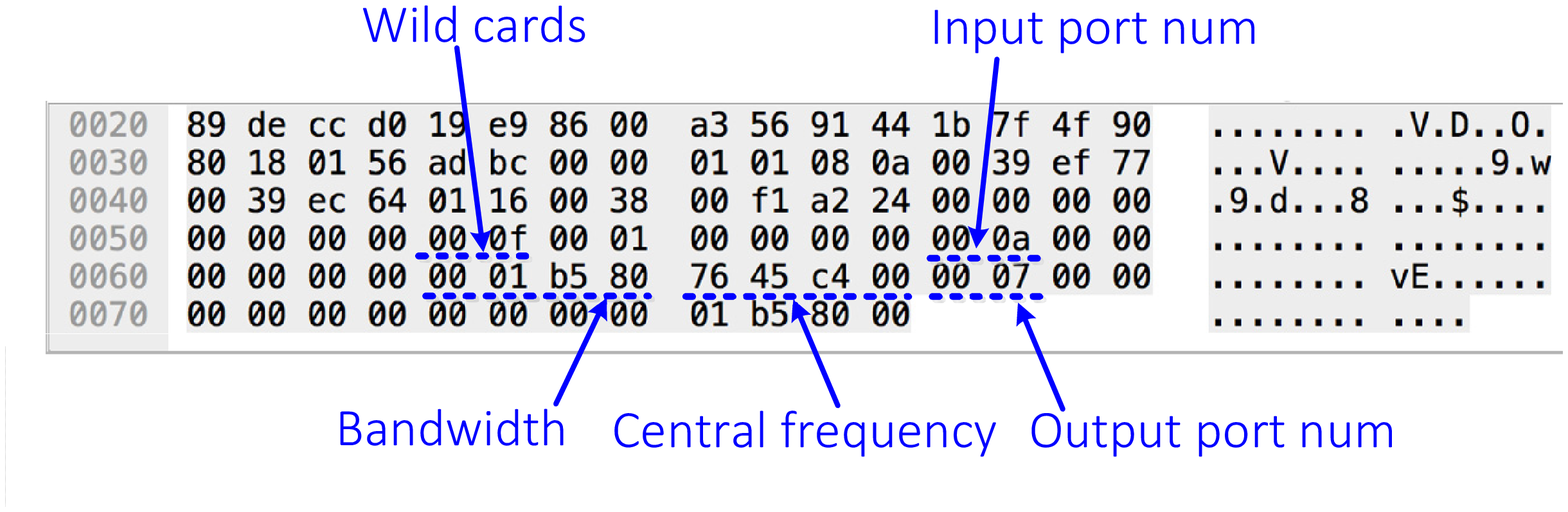}
\vspace*{-10pt}
\caption{Real-time monitoring information: OpenFlow message for the configuration of the WSS-1.}
\label{fig:OF_2ndWSS}
\end{figure}

All the switch and V-BVT re-configurations described in the above paragraphs are conducted using OF messages through ODL controllers shown in Fig. \ref{fig:OF_Ethernet} and Fig. \ref{fig:OF_2ndWSS}. In Fig. \ref{fig:OF_Ethernet}, the frame coloured in black represents the OSNR monitoring information obtained by the WA and are fetched by the V-BVT application running on top of the controller. The frame in red indicates the creation of two new transmission services, each of which has different data rate. The transmission in-/out-put ports in the Ethernet switch, the subcarriers and modulation formats are selected for accommodating these new services. The frame coloured in blue indicates the moment when the monitored traffic at the in-put ports of these services drops below the threshold that their sum is smaller than a threshold. Finally, the frame in green indicates the re-configuration results based on the new optimised decision made by the V-BVT application.  

% \begin{figure}[h!]
 % % \vspace*{10pt}
% \includegraphics[width=0.8\textwidth]{OF_3rdopof}
% % \vspace*{10pt}
% \caption{Real-time monitoring information: OpenFlow message for configuring WSS}
% \label{fig:OF_3rdWSS-1}
% \end{figure}

Fig. \ref{fig:OF_2ndWSS} indicates the OF messages when configuring the 4x16 WSSs within the V-BVT. This message is captured by the network protocol analyser software Wireshark. It uses this format of Hexadecimal numbers to show our WSS configuration values within the OF flow\_mod message.

Similar to Fig. \ref{fig:OF_Ethernet}, in order to create virtual transceivers for accommodating the groomed traffic service, the input port A is configured for WSS to allow the input of the subcarrier pool. This configuration can be seen from the OF message as a Hexadecimal number ``00 0a", indicating the input port number A is set among the WSS's four input ports labelled as A, B, C and D. Similarly, ``00 07" equals 7 in decimal, indicting the output port number 7 is selected among the WSS's 16 output ports. Therefore, the switching between ports A and 16 is enabled. To configure a channel that to be switched via this connection, the central frequency of the channel is specified at these two ports as 198.4283648 GHz, which can be seen from another Hexadecimal number ``76 45 c4 00". Apart from the central frequency, the filtering width of these two ports A and 16 is also configured as 112.0 GHz, showing as ``00 01 b5 80" in the OF message. This will allow the switching of the channel(s) centred at 198.4283648 GHz with a bandwidth smaller than 112.0 GHz between A and 16.

\section{Conclusion} \label{sec6:conclusion}
In this paper, for the first time, we have experimentally demonstrated a proposed optical virtualisation scheme utilizing the joint technologies of V-BVTs and real-time monitoring in both the optical and transport networks. This scheme achieves optimisation in V-BVT optical infrastructures and reconfiguration of Ethernet switch resources through an OpenDayLight controller, during on-demand creation of virtual transceivers. The experiment also reflects the feasibility of using multi-function monitoring to facilitate a holistic optical network virtualisation. It also indicates the necessity of adopting V-BVT architecture in the hardware level as part of the software defined optical network to enable the network resource (re-)confirguation and support the network virtualisation in an efficient manner.

% or


% \appendices

% \section{}

\section*{Acknowledgment}
This work is supported by EPSRC grant EP/L020009/1: Towards Ultimate Convergence of All Networks (TOUCAN).

\ifCLASSOPTIONcaptionsoff
  \newpage
\fi



%\bibliographystyle{IEEEtran}

% \end{thebibliography}

\bibliographystyle{IEEEtran}
\bibliography{References/monitoring2016_bib} % Path to your References.bib file


% \end{IEEEbiography}


% % insert where needed to balance the two columns on the last page with
% %\newpage



%\vfill


\end{document}